# A Deep Reinforcement Learning Approach for Trading Optimization in the Forex Market with Multi-Agent Asynchronous Distribution

Davoud Sarani, Dr. Parviz Rashidi-Khazaee

*Abstract*— **In today's forex market traders increasingly turn to algorithmic trading, leveraging computers to seek more profits. Deep learning techniques as cutting-edge advancements in machine learning, capable of identifying patterns in financial data. Traders utilize these patterns to execute more effective trades, adhering to algorithmic trading rules. Deep reinforcement learning methods (DRL), by directly executing trades based on identified patterns and assessing their profitability, offer advantages over traditional DL approaches. This research pioneers the application of a multi-agent (MA) RL framework with the state-of-the-art Asynchronous Advantage Actor-Critic (A3C) algorithm. The proposed method employs parallel learning across multiple asynchronous workers, each specialized in trading across multiple currency pairs to explore the potential for nuanced strategies tailored to different market conditions and currency pairs. Two different A3C with lock and without lock MA model was proposed and trained on single currency and multi-currency. The results indicate that both model outperform on Proximal Policy Optimization model. A3C with lock outperforms other in single currency training scenario and A3C without Lock outperforms other in multi-currency scenario. The findings demonstrate that this approach facilitates broader and faster exploration of different currency pairs, significantly enhancing trading returns. Additionally, the agent can learn a more profitable trading strategy in a shorter time.**

## I. INTRODUCTION

The Forex market is the largest financial market in the world for currency trading due to its high daily trading volume [1]. Financial trading can be either manual or algorithmic. Manual trading relies on technical and fundamental analysis, while algorithmic trading is carried out by computers using rule-based or machine learning (ML)-based strategies [2]. Humans may make erroneous trading decisions due to their emotional and psychological factors. Therefore, traditional trading can be susceptible to human errors, while computers are not adversely affected by these factors and are more precise in executing trading decisions compared to humans [2]. The rule-based approach in algorithmic trading involves identifying profitable trading signals from specific market movements by analyzing the time series of price data. The ML approach, on the other hand, automatically learns patterns that lead to predictable market movements and generates trading signals accordingly [3].

The ML algorithms can extract patterns and relationships from time series data without the need for predefined instructions or strategies set by domain experts, discovering profitable trading patterns that may not be discernible to humans [2]. In recent years, ML, serving as an intelligent agent, has replaced traditional human decision-making methods. This transition, particularly with the advent of deep learning (DL), has enhanced algorithm performance. These capabilities encompass the extraction of complex patterns from data and the immunity to emotions that could impact their performance [4]. The ML approach to algorithmic trading can be divided into supervised learning and RL approaches [5]. The supervised learning approach focuses on predicting stock prices or price trends in the next time step [5]. One drawback of supervised learning and DL algorithms for prediction and classification is the absence of executing trading decisions [1]. The predicted prices or trends cannot be directly associated with the order actions. Therefore, after price prediction, prior knowledge of the financial domain is required for selecting trading actions, and the accuracy of the prediction influences the decision-making process [5]. Price prediction aims to construct a model capable of forecasting future prices. However, the algorithmic trading extends beyond price prediction, focusing on active participation in the financial market (such as selecting trading positions and the number of shares traded) to maximize profits [4]. Supervised learning methods have shown significant potential for predicting financial markets, such as the Forex market. However, their prediction accuracy may not be sufficient for algorithmic applications in real markets due to the fluctuations, instability, and uncertain nature of financial time series data [2]. Financial data is noisy, and this might be a reason why supervised learning methods have not been successful in the past [6].

Supervised learning is not suitable for problems involving

This paragraph of the first footnote will contain the date on which you submitted your paper for review, which is populated by IEEE. It is IEEE style to display support information, including sponsor and financial support acknowledgment, here and not in an acknowledgment section at the end of the article. For example, "This work was supported in part by the U.S. Department of Commerce under Grant 123456." The name of the corresponding author appears after the financial information, e.g.

\* -Correspond author: Dr. Parviz Rashidi-Khazaee.

Davoud Sarani, is Mastery program student at Information Technology and Computer Engineering Department, Urmia University of Technology, Urmia, Iran (e-mail: sarani.davoud@it.uut.ac.ir).

Dr. Parviz Rashidi-Khazaee is Assistant Professor at Information Technology and Computer Engineering Department, Urmia University of Technology, 4km Band Road, Urmia, Iran (e-mail: p.rashidi@uut.ac.ir).

long-term and delayed rewards, such as trading in financial markets. For addressing decision-making issues (conducting trades) in an uncertain environment (financial market), RL is a more appropriate choice [4]. RL does not require supervision labels. In RL, an agent interacts with the environment and receives rewards or penalties. In a financial trading environment, the agent decides what trading actions to take and is rewarded or penalized based on its trading performance [3]. Training an RL agent eliminates the complexity of manual label selection and allows it to determine which trading positions have predictable results and value based on the received rewards (trading profits). It enables the direct optimization of the profitability and loss-related metrics [7]. The newly developed Deep Reinforcement Learning (DRL) algorithms can independently make optimal decisions in complex environments and perform better than basic strategies [2]. It has been shown that the DRL algorithms, which use the potential advantage of RL-based trading strategies, outperform rule-based strategies [2].

DRL algorithms are divided into two single agents and multi-agent methods. Carapuco et al. used the Q-learning algorithm and a method for more efficient use of available historical tick data of the EUR/USD pair including bid and spread to improve training quality and showed a stable equity growth and minimal drawdowns [8]. Tsantekidis et al. apply the single agent learning method using the Proximal Policy Optimization (PPO) algorithm to Forex trading and propose a market-wide training approach with a feature extraction method to enable agents to adapt to diverse currency pairs [3].

The complexity and dynamic nature of the financial market makes it necessary to find an optimal trading strategy, prompting the exploration of multi-agent systems within DRL, which generally outperforms single-agent approaches [9]. In the Multi-agent domain, Shavandi and Khedmati propose a hierarchical DRL framework for forex trading, specialized in various periods [2]. These independent agents communicate through a hierarchical mechanism, aggregating the intelligence across different timeframes to resist noise in financial data and improve trading decisions. Korczak and Hernes integrate DL with a multi-agent system to enhance the ability to generate profitable trading strategies, employing a supervisory agent to orchestrate diverse trading strategies and select the most promising recommendations [10]. Ma et al. underscores the importance of multi-agent systems in portfolio management, showcasing superior performance compared to single-agent strategies [9]. Tsantekidis et al. underscore the effectiveness of knowledge distillation from multiple teacher agents to student agents, thereby enhancing the trading performance of students. It emphasizes the significance of diversifying teacher models to trade various currencies in volatile markets, thus improving the performance of students [7].

Parallel multi-agent algorithms like asynchronous advantage actor-critic (A3C), IMPALA, and SeedRL could be used in the forex trading market. From these categories, A3C plays an important role in the forex market. Li et al. employ A3C algorithms to tackle feature extraction and strategy adaptation issues and showcase superior performance in stock and futures markets [11]. Kang et al. implement the A3C algorithm for stock selection and portfolio management, observing enhanced stability and convergence in training but encountering less impressive performance during testing, possibly due to data limitations and neural network (NN) simplicity [12]. Ponomarev et al. investigate the A3C algorithm's efficacy in algorithmic trading, particularly on Russia Trading System (RTS) Index futures, by creating a trading environment, testing NN architectures, and analyzing historical data, underscoring the algorithm's profitability [13]. The implementation of parallel workers with workload distribution in the A3C [14]. enhances computational efficiency, reduces agent training time, and effectively explores the environment, learning an improved optimal policy in less time. Through parallel environment exploration, A3C outperforms other algorithms in terms of diverse experiences and trading profitability [13].

While the use of multiple agents to make the ultimate trading decision has shown considerable advancements [2, 7, 10] and is technically feasible to train agents to handle market-wide trading on a range of currency pairs [3, 7], training multiple teachers agents and distilling trading decisions to student agents enhanced the trading performance of the student, using a diverse subset of currency pairs to train teachers can improve the student proficiency [7]. Training DL models is time-consuming, but implementing distributed computing can expedite the learning process [10]. Thus distilling only profitable trading decisions to students [7]. limits the student knowledge and shortcoming the overall acknowledgment of various circumstances. Therefore, not training in a distributed manner leads to suboptimal resource utilization.

So far, the A3C algorithm has not been utilized for parallel training multiple agents across various currency pairs to share their knowledge with each other and develop a generalized optimal policy for the Forex market. In this study, we aim to pioneer this approach and explore its effectiveness. Additionally, another objective of this work is to compare single-agent (SA) with multi-agent (MA) approaches. The key contribution of this study is to utilize distributed training to develop an agent capable of trading diverse pairs in financial markets like forex. This approach aims to enhance agent learning and policy generalization across various market conditions, improve exploration efficiency, and accelerate learning and exploration in different environmental segments, thereby enabling the acquisition of more robust and generalized policies. The utilization of multiple parallel workers with the A3C algorithm, enables the acquisition of extensive experience in diverse environments, resulting in quicker adaptation and convergence to abrupt changes within financial markets.

The forthcoming paper is structured as follows: Section II offers a comprehensive review of previous works relevant to multi-agent RL. Section III explores the details of RL models, while Section IV discusses our methodology. Following this, Section V provides an overview of the implementation details of the method. Section VI analyzes and interprets the results.

Finally, section VII provides conclusions.

## II. RELATED WORKS

Carapuco et al. employ the Q-learning algorithm and a method for more efficient utilization of historical tick data of the EUR/USD pair, including bid price and spread, to enhance training quality. They demonstrate stable equity growth and minimal drawdowns. Despite the non-deterministic and noisy nature of financial markets, the study showcases stable learning in the training dataset and the Q-network's ability to identify relationships in financial data, resulting in profitable trading in a test dataset. The potential for further optimization in parameters, network topology, and model selection methods suggests promising avenues for future exploration in algorithmic trading strategies. The study proposes future work to concentrate on optimizing parameters, network topology, and model selection methods, alongside exploring enhancements in dataset selection and financial optimizations [8].

Shavandi and Khedmati introduce a novel DRL multi-agent framework tailored for financial trading, where agents specialize in specific time periods. These agents operate independently yet collaborate through a hierarchical feedback mechanism, facilitating the transmission of knowledge from higher timeframe agents to lower ones. This mechanism serves to resist noise in financial data, enabling the aggregation of intelligence across different timeframes. By sharing insights and learning characteristics within each interval, the framework outperforms both independent agents and rule-based strategies. Its primary objective is to establish intertemporal learning interactions via collective intelligence among multiple agents, enabling adaptation to noise and utilization of price movement details for enhanced trading performance [2].

The study of Korczak and Hernes discusses the integration of DL into a multi-agent framework for creating profitable trading strategies in Forex. The system, called A-Trader, utilizes trading actions and fuzzy logic to make decisions based on factors like confidence levels and probabilities. A supervisory agent oversees decision-making, coordinating various trading strategies and selecting the most appropriate suggestions for investment decisions. DL is employed to forecast financial data, aiming to enhance A-Trader's ability to offer profitable trading recommendations. However, a drawback of DL is its time-consuming learning mode, which could be mitigated by employing distributed cloud computing [10].

Ma et al. introduce a novel approach to financial portfolio management, leveraging a multi-agent DRL algorithm with trend consistency regularization to recognize consistency in stock trends, guiding the agent's trading strategies. this approach divides stock trends into two categories and trains two agents with the same policy model and value model and different reward functions are constructed, differing in regularization, and enhanced adaptability to market conditions. By dynamically switching between agents based on market conditions, the proposed algorithm optimizes portfolio allocation, achieving higher returns and lower risk compared to existing algorithms [9].

A knowledge distillation method proposed by Tsantekidis Et al. for training RL agents in the financial market by employing teacher agents in diverse sub-environments to diversify their learned policies. Subsequently, student agents then utilize profitable knowledge from these teachers to emulate existing trading strategies. It emphasizes that diversifying teacher models for trading various currencies and knowledge distillation from multiple teacher agents can significantly enhance the performance of students in volatile financial markets and for this purpose, pre-processed observations of past candlestick price patterns to identify percentage differences between sampled prices. They also suggest that using the Policy Gradient approach is more efficient than the DQN approach [7].

Tsantekidis et al. also a suggests reward-shaping method based on prices for Forex trading with a DRL approach using the Proximal Policy Optimization (PPO) algorithm. This approach enhances agent performance in terms of profit, Sharpe ratio, and maximum drawdown. The authors also employ a data preprocessing and fixed-feature extraction method to enable agent training on various Forex currency pairs, facilitating the development of RL agents across the wide range of pairs in the market while mitigating overfitting. The paper emphasizes that RL agents have typically been trained to trade individual assets, whereas human traders can adapt their trading strategies to different assets and conditions. To overcome this limitation, they propose a market-wide training approach that extracts valuable insights from various financial instruments. The proposed feature extraction method enhanced the effective data processing from diverse distributions [3].

Li et al. propose a framework for algorithmic trading by utilizing DQN and A3C algorithms with SDAEs and LSTM networks, to address feature extraction and strategy adaptation challenges. They are resulting in superior performance compared to baseline methods in both stock and futures markets, demonstrating substantial improvement and potential for practical trading applications. Specifically, the SDAEs-LSTM A3C model learns a more valuable strategy and surpasses LSTM in predictive accuracy [11].

Kang et al. apply the A3C algorithm to stock selection and portfolio management using a subset of S&P500 index stocks, training asynchronously, with multiple environments initiated at different times to simulate experience buffers. Despite notable improvements in stability and convergence during training (shortening the training process and speeding up the convergence process), the model's performance during the test period is not as impressive, potentially due to limitations in data availability and the simplicity of the NN architecture. recommended incorporating more data and features to boost performance, emphasizing the significance of robust model architecture and adequate data for effective results [12].

Ponomarev et al. explore the A3C algorithm, in algorithmic trading, focusing on trading RTS Index futures. The study constructed a trading environment, experimenting with various NN architectures, testing on historical data and highlighting the potential profitability and attractiveness of the algorithm for investment, and verifying the effectiveness of Long short-term memory (LSTM) and dropout layers, while debating the impact of a reward function and the number of neurons in hidden layers.

emphasizing the importance of optimizing architectures for real trading systems [13].

## III. REINFORCEMENT LEARNING (RL) MODELS

In this research, algorithms based on PPO and A3C were utilized to train an RL agent capable of executing trades in the forex market.

### A. Actor-Critic

Actor-Critic employs an NN architecture with two components: an actor and a critic; The actor learns a policy (the strategy for selecting actions), while the critic network estimates the expected future reward and reduces variance by providing a baseline for advantage estimates. The combination of actor-critic architecture helps stabilize training by reducing the high variance typically associated with policy-based methods.

### B. Advantage function

The advantage function is used to compute the policy gradient. It guides the actor to select actions that lead to better outcomes and addresses the credit assignment problem by providing feedback on the quality of chosen actions. The advantage is calculated as (1), where the Q-value (action value) represents the expected cumulative reward by taking a specific action in a particular state and then following a certain policy thereafter. The V-value (critic value) represents the expected cumulative reward that can be obtained from a particular state onwards, following a certain policy.

$$A(sa) = Q(sa) - V(s) \qquad (1)$$

### C. Proximal Policy Optimization (PPO)

The PPO [15] is an on-policy RL algorithm that optimizes the agent's policy to maximize the expected cumulative reward. It works by iteratively collecting data through interactions with the environment and updating the policy to improve its performance. The policy maintains a probability distribution over actions for each state, represented by a NN. This algorithm computes the surrogate objective to guide policy gradient for actions with higher returns, and the clipped objective to limit policy updates to maintain stable training and prevent learning disruptions. By constraining policy updates and maintaining a trust region between old and new policies, it addresses issues of high variance and unstable learning, ensuring that the policy update does not deviate too far from the current policy, preventing major learning disruptions; and balances exploration and exploitation by iteratively collecting data with the current policy and optimizing the policy using that data. The loss in PPO is calculated as (2), in this equation the $\text{clip}(x, a, b)$ is a clipping function that clips the value $x$ to the range $[a, b]$ and $\mathcal{H}(\pi_\theta(\cdot|s_t))$ represents the entropy of the policy, the $r_t(\theta)$ is the ratio of the probability of the new policy to the old policy and calculated as (3).

$$L(\theta) = \mathbb{E}_t\left[\min\left(r_t(\theta)\hat{A}_t, \text{clip}(r_t(\theta), 1-\epsilon, 1+\epsilon)\hat{A}_t\right) - \beta\mathcal{H}(\pi_\theta(\cdot|s_t))\right] \qquad (2)$$

$$r_t(\theta) = \frac{\pi_\theta(a_t|s_t)}{\pi_{\theta_{\text{old}}}(a_t|s_t)} \qquad (3)$$

### D. Asynchronous Advantage Actor-Critic (A3C)

The A3C is an advanced variant of the actor-critic architecture, The advantage in A3C refers to the advantage function, which quantifies how much better or worse a particular action is compared to the average action [14]. In A3C, Multiple local workers run in parallel with their copy of the policy network and environment, collecting experiences and updating the global networks asynchronously. This enables efficient utilization of resources, faster convergence, better exploration, and more sample-efficient learning.

Accumulate gradients with respect to the local policy network parameters $\theta'$ using the policy gradient and advantage estimation are calculated as $\nabla_{\theta'}\log\pi(a_i|s_i;\theta')(R - V(s_i;\theta'_v))$, and the accumulated gradients with respect to the local value network parameters $\theta'_v$ using the squared temporal difference error is computed as $\partial(R - V(S_i;\theta'_v))^2/\partial\theta'_v$, Where $(R_t - V(s_t))$ is the difference between the estimated value $V(s_t)$ and the observed reward $R_t$. The $\partial/\partial\theta'_v$ is the partial derivative with respect to the parameter $\theta'_v$ and it is used to find how a small change in the parameter $\theta'_v$ affects this expression.

The R is the total discounted return calculated as $R \leftarrow r_i + \gamma R$, in this expression $r_i$ is an immediate reward received at the time step $i$, and $\gamma$ is the discount factor to discount the values of future rewards.

The Policy loss (actor loss) for the N time steps in A3C is computed as (4), The Value loss (critic loss) in A3C guides the value function towards better approximations of the expected return (R) and it is the mean squared error between the estimated value function(V) and the actual return. The value loss for the N time steps is computed as (5).

$$L_{\text{policy}} = -\frac{1}{N}\sum_{t=1}^{N} \log\pi(a_t|s_t) \cdot A_t \qquad (4)$$

$$L_{\text{critic}} = \frac{1}{N}\sum_{i=1}^{N} (R_i - V(s_i;\theta_v))^2 \qquad (5)$$

In the asynchronous updating process of the global worker from the local workers, the global network aggregates gradients from multiple local workers and updates its parameters, while each local worker updates its parameters independently based on its local experiences. Assuming the $\theta$ and $\theta_v$ are the shared parameters of the global worker, $\theta'$ and $\theta'_v$ represent the parameters of local workers; based on Algorithm S3 [14], the asynchronous update for the actor can be defined as (6) and (7) for the critic.

$$\theta': d\theta \leftarrow d\theta + \nabla_{\theta'}\log\pi(a_i|s_i;\theta')(R - V(s_i;\theta'_v)) \qquad (6)$$

$$\theta'_v: d\theta_v \leftarrow d\theta_v + \frac{\partial(R - V(s_i;\theta'_v))^2}{\partial\theta'_v} \qquad (7)$$

## IV. METHODOLOGY

In this section, we will first address the necessary prerequisites. Then we will examine and discuss our proposed method for enhancing the speed of the agent and learning optimal policies in the forex market.

*A. Data Preparation*

Candlestick data of different currency pairs in the Forex market in one-hour timeframes have been used to train the agents. This data is directly retrieved from the forex broker's terminal.

Due to price fluctuations in various currency pairs and the recurring patterns in candlestick models, using raw candlestick data is inefficient. To address this issue and normalize the data, the method described in [3] is employed (8). This method utilizes the ratio of candlestick changes from time series data to create 5 new features as (9) instead of using raw data.

$$\begin{aligned} x1_t &= \frac{Pc_t - Pc_{t-1}}{Pc_{t-1}} \\ x2_t &= \frac{Ph_t - Ph_{t-1}}{Ph_{t-1}} \\ x3_t &= \frac{Pl_t - Pl_{t-1}}{Pl_{t-1}} \\ x4_t &= \frac{Ph_t - Pc_t}{Pc_t} \\ x5_t &= \frac{Pc_t - Pl_t}{Pc_t} \end{aligned} \tag{8}$$

$$X_t = [x1_t, x2_t, x3_t, x4_t, x5_t] \tag{9}$$

$$X = [X_{t-window}, \ldots, X_{t-2}, X_{t-1}, X_t] \tag{10}$$

*B. Reward Function*

The evaluation of agent actions is carried out using rewards received from the environment. The agent through its interactions with the environment and the selection of a trading decision (12), receives rewards based on the decision it made (13), which can be either profitable (positive reward) or detrimental (negative reward). The received reward is a normalized value, determined by the price changes in consecutive candlestick closing prices within two successive time intervals (11).

$$z_t = \frac{Pc_t - Pc_{t-1}}{Pc_{t-1}} \tag{11}$$

$$\delta_t \in \begin{cases} 1 & \text{Long} \\ -1 & \text{Short} \\ 0 & \text{Neutral} \end{cases} \tag{12}$$

$$r_t = \delta_t * z_t \tag{13}$$

The value of δ is 1 for a long trading decision, -1 for a short trading decision, and 0 when closing a trade or staying out of the market.

In this work, the primary objective was to highlight training time efficiency and the learning of an optimal policy, and therefore, the commission is not taken into the equation. Also, it is assumed that there is no spread thus the bid and ask prices are equal and rewards are calculated based on the closing price.

*C. Proposed Reinforcement Learning (RL) Model*

RL is comprised of the interaction between two main components: the agent and the environment. The environment's function involves both preparing financial market data as the observation space and interacting with the received trading decisions as the action space. The observation space consists of historical financial data as (10) in conjunction with trading decisions made within the specified past window timeframe which are categorized into one of three groups: opening a long trading position, opening a short trading position, closing an existing position, or staying out of the market. In RL, the agent receives an observation state, makes a trading decision, and subsequently receives a reward from the environment based on profit or loss.

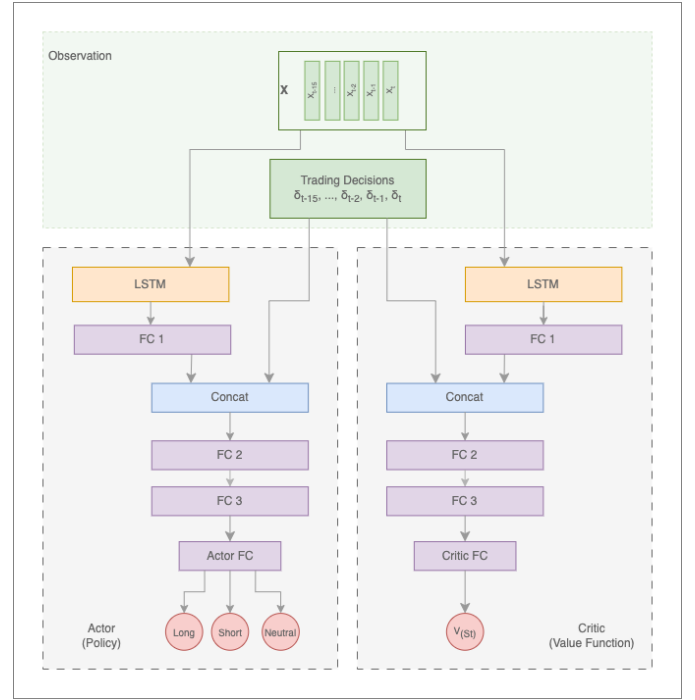

**Fig. 1.** The proposed Actor-Critic model.

Figure 1 and Scheme 1 showed the structure of proposed model. A uniform NN model was used for both algorithms. The model's input consists of two parts: the first part contains an LSTM layer for normalized time series data (10), and the second part is a linear layer for the one-hot encoded trading decisions made within a chosen time interval of a specific window size.

The number of neurons in the hidden LSTM layer is set to 128, The output from the LSTM is first passed through a layer consisting of 32 neurons. Then, it is combined with the trading decisions from previous steps, based on window size, and fed into a fully connected (FC) layer.

The output from the previous steps is combined and sent to a FC layer with 64 neurons. This FC layer then sends the output to either an actor or critic FC layer, depending on whether it's for trading decisions or evaluation. The actor FC layer's output represents the number of trading decisions made, while the critic FC layer's output is used to evaluate the actor's performance.

```
ActorCritic(
  (actor): LSTM_MLP(
    (lstm): LSTM(5, 128, batch_first=True)
    (fc1): Linear(in_features=128, out_features=32, bias=True)
    (fc2): Linear(in_features=80, out_features=64, bias=True)
    (fc3): Linear(in_features=64, out_features=64, bias=True)
    (output_layer): Linear(in_features=64, out_features=3, bias=True)
  )
  (critic): LSTM_MLP(
    (lstm): LSTM(5, 128, batch_first=True)
    (fc1): Linear(in_features=128, out_features=32, bias=True)
    (fc2): Linear(in_features=80, out_features=64, bias=True)
    (fc3): Linear(in_features=64, out_features=64, bias=True)
    (output_layer): Linear(in_features=64, out_features=1, bias=True)
  )
)
```

Scheme 1 - Scheme of The Actor-Critic Model

## V. EXPERIMENTS

The proposed methods are examined in this section, and their results are compared. Initially, the dataset is reviewed, followed by a discussion of the training methods and parameters. Finally, the impact of asynchronous learning using multiple agents in parallel is investigated.

### A. Training and evaluation Dataset

A dataset based on transaction data from the Forex market has been used to train and evaluate algorithms. The datasets were obtained and saved using the MetaTrader5 terminal and contain price movement charts (candles) for major, minors, and cross-currency pairs within a one-hour timeframe from 2009 to mid-2017 The data was divided into two parts. The data from 2009 to the end of 2016 was used for training, and the first four-month data from 2017 were used for back-testing.

### B. Evaluation Metrics

The models are evaluated using key measures such as "Return," which indicates whether a strategy is profitable or not. The "Sharpe Ratio" helps to assess risk and returns, while the "Profit Factor" measures the amount of money made versus lost. "Maximum Drawdown" reveals the strategy's largest loss. These measures are utilized to analyze and compare models and determine their relative performance.

$$Return = \frac{(FinalPrice - InitialPrice)}{InitialPrice} \times 100 \quad (14)$$

$$SharpeRatio = \frac{R_p}{\sigma_p} \quad (15)$$

In the Sharpe Ratio, the $R_p$ is the mean returns, and $\sigma_p$ is the standard deviation of the returns.

### C. Parameters

The parameters used in all experiments were as follows: a discount factor of 0.99, a learning rate of 0.00004, and a time window of 16, which considers previous time series samples as inputs. In addition, the same reward function was employed to evaluate the performance of all models, and the same seed was used for both model training and evaluation.

### D. Training

Training process for both single-agent (SA) and multi-agent (MA) approaches, was conducted using two different scenarios: single-currency (SC) and multi-currency (MC). In the SC scenario, the EUR/USD currency pair was used for training, with a randomly chosen starting point for each training episode. In the MC scenario, the training was carried out on 28 different currency pairs, where at each episode, a random currency pair and starting point were selected. In both approaches, each training episode consisted of 600 steps, with a total of one million steps for each training approach.

The PPO algorithm has been used for SA and the A3C algorithm has been used for the MA approach.

### E. Single-agent training (SA)

In SA training at each training step, 5 extracted features at the time $t$ along with the 15 previous extracted features resulting in a total of 16-row, are fed into the LSTM layer, and the output, along with selected trading actions at each step in a one-hot encoding shape, are passed to the next corresponding FC layers.

TABLE I
SC SCENARIO EVALUATION RESULTS

| Model | Return | Sharpe Ratio | Win Rate | Profit Factor |
|---|---|---|---|---|
| SA | 71.27 % | 1.27 | 51.80 % | 1.20 |
| MA-Lock | 98.22 % | 1.46 | 51.75 % | 1.27 |
| MA-NoLock | 81.33 % | 1.58 | 52.09 % | 1.20 |

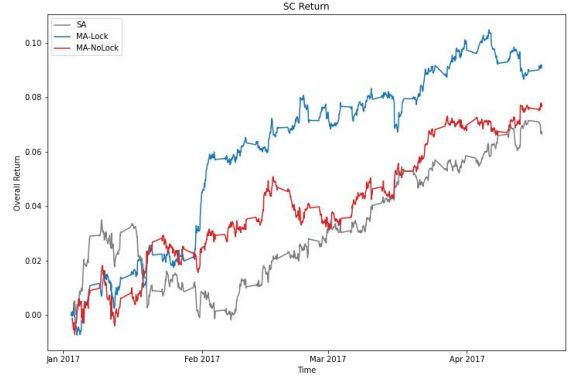

**Fig. 2.** Comparative return of backtesting in SC scenario.

### F. Multi-Agent training (MA)

The MA training process follows the same parameters and conditions as SA. Five local workers are employed for parallel training to speed up the learning process. Each local worker undergoes 20 training steps before updating the global worker. This process continues until the end of one episode, after which a new episode begins.

During the training with A3C, two different approaches were examined. In the MA-Lock, a lock mechanism was implemented, which allowed only one local worker at a time to make updates to the global worker. the MA-NoLock did not have a locking mechanism and multiple local workers could interact with the environment and concurrently share updates with the global worker without any conflicts. Also, an optimizer with shared parameters was implemented across all workers

## V. RESULTS

The performance evaluation for each approach was conducted using the "backtesting" [16] library over four months with test data. In the A3C algorithm, the global worker is used for evaluation. In backtesting, when the agent repeatedly selects the same trading decisions in multiple consecutive environmental states, the chosen trading decisions are regarded as one continuous trade.

### A. Single Currency Pair scenario (SC)

When comparing the performance of the SA and MA in a SC training scenario, both MA models, with and without a lock mechanism, demonstrated higher returns, more favorable risk-adjusted returns (as indicated by the Sharpe Ratio), and a slightly better profit factor compared to SA (Table I, Fig. 2).

*B. Multi-currency pairs scenario*

In the MC scenario, training the SA maintained relatively consistent yet generally lower returns, while MA exhibited mixed performance in returns when comparing MA-Lock and MA-NoLock mechanisms (Table III).

TABLE V
MC BACKTESTING DRAWDOWN ON 9 PAIRS

| Model | Pair | Max. Drawdown | Avg. Drawdown | Max. Drawdown Duration | Avg. Drawdown Duration |
|---|---|---|---|---|---|
| SA | EURUSD | -14.52 % | -2.74 % | 17 days 03:00:00 | 2 days 01:00:00 |
| | AUDUSD | -22.82 % | -4.43 % | 55 days 07:00:00 | 6 days 03:00:00 |
| | EURGBP | -20.93 % | -2.31 % | 24 days 04:00:00 | 2 days 03:00:00 |
| | AUDCAD | -21.18 % | -4.20 % | 49 days 13:00:00 | 4 days 15:00:00 |
| | EURCHF | -28.46 % | -10.62 % | 70 days 02:00:00 | 26 days 05:00:00 |
| | EURAUD | -26.24 % | -3.18 % | 63 days 01:00:00 | 3 days 05:00:00 |
| | USDCAD | -25.03 % | -6.25 % | 76 days 03:00:00 | 6 days 23:00:00 |
| | GBPNZD | -47.93 % | -14.56 % | 55 days 03:00:00 | 9 days 06:00:00 |
| | GBPUSD | -37.32 % | -5.55 % | 90 days 13:00:00 | 6 days 10:00:00 |
| MA-Lock | EURUSD | -15.66 % | -1.82 % | 36 days 12:00:00 | 1 days 17:00:00 |
| | AUDUSD | -48.87 % | -3.73 % | 94 days 10:00:00 | 5 days 11:00:00 |
| | EURGBP | -27.32 % | -3.11 % | 101 days 23:00:00 | 9 days 11:00:00 |
| | AUDCAD | -50.37 % | -4.47 % | 94 days 12:00:00 | 6 days 12:00:00 |
| | EURCHF | -12.81 % | -1.59 % | 34 days 19:00:00 | 2 days 15:00:00 |
| | EURAUD | -49.62 % | -4.31 % | 84 days | 4 days 16:00:00 |
| | USDCAD | -40.44 % | -10.89 % | 82 days 09:00:00 | 14 days 21:00:00 |
| | GBPNZD | -60.26 % | -6.12 % | 38 days 17:00:00 | 3 days 22:00:00 |
| | GBPUSD | -16.10 % | -3.13 % | 15 days 13:00:00 | 2 days 03:00:00 |
| MA-NoLock | EURUSD | -20.05 % | -2.36 % | 31 days 01:00:00 | 1 days 21:00:00 |
| | AUDUSD | -11.45 % | -1.55 % | 20 days 03:00:00 | 1 days 09:00:00 |
| | EURGBP | -33.52 % | -4.28 % | 48 days 10:00:00 | 3 days 20:00:00 |
| | AUDCAD | -14.56 % | -2.82 % | 46 days 22:00:00 | 2 days 20:00:00 |
| | EURCHF | -8.01 % | -1.30 % | 17 days 08:00:00 | 1 days 12:00:00 |
| | EURAUD | -45.69 % | -7.48 % | 65 days 16:00:00 | 5 days 19:00:00 |
| | USDCAD | -66.68 % | -16.59 % | 83 days 02:00:00 | 14 days 23:00:00 |
| | GBPNZD | -61.36 % | -12.65 % | 65 days 09:00:00 | 7 days 08:00:00 |
| | GBPUSD | -18.08 % | -2.76 % | 27 days 22:00:00 | 1 days 12:00:00 |

MA-NoLock stood out with the highest average return (Table IV, Fig. 3), favorable risk-adjusted returns (positive Sharpe Ratio), and the best profit factor. On the other hand, SA had the lowest return, with slightly lower risk-adjusted returns and profit factors. MA-Lock fell in between these two, displaying slightly better risk-adjusted returns compared to SA but still lagging behind MA-NoLock in terms of return and profit factor.

Notably, MA-NoLock has the advantage of recovering from drawdowns in a shorter time compared to the other two entities, despite not having the lowest drawdown magnitude. This suggests that MA-NoLock has a more efficient and faster recovery mechanism or risk management strategy in place. While SA has a lower drawdown magnitude, its longer drawdown duration indicates that it takes more time to bounce back from losses. This observation underscores the importance of not only considering the depth of drawdowns but also the time it takes to recover when evaluating trading models (Table V, Table VI).

TABLE VI
MC BACKTESTING average DRAWDOWN on 9 pairs

| Model | Max. Drawdown | Avg. Drawdown | Max. Drawdown Duration | Avg. Drawdown Duration |
|---|---|---|---|---|
| SA | -27.16% | -5.98% | 53 days, 10 hours | 7 days, 5 hours |
| MA-Lock | -35.72% | -4.35% | 60 days, 6 hours | 5 days, 5.78 hours. |
| MA-NoLock | -31.04% | -5.76% | 37 days, 18.33 hours | 4 days, 11.56 hours |

TABLE VII
MC SCENARIOS BACKTESTING evaluation results on EUR/USD

| Model | Return | Sharpe Ratio | Win Rate | Profit Factor |
|---|---|---|---|---|
| SA | 54.21% | 0.78 | 51.61 | 1.05 |
| MA-Lock | 73.72% | 1.72 | 52.64 | 1.21 |
| MA-NoLock | 70.12% | 1.29 | 49.16 | 1.19 |

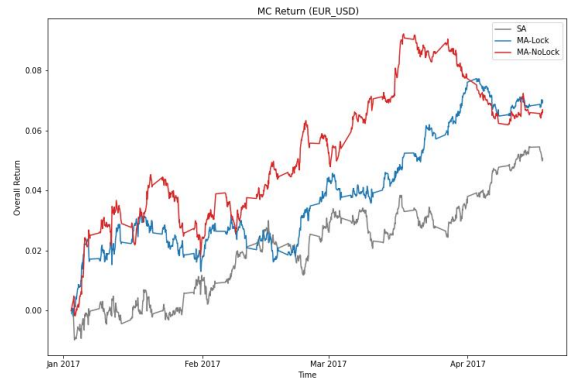

**Fig. 4.** Comparative MC scenarios backtesting results on EUR/USD.

In The MC scenario, the backtesting results on EUR/USD showed that both MA models outperformed the base SA model (Table VII, Fig. 4). Additionally, the MA-Lock performed better than the MA-NoLock, highlighting the superiority of using the A3C with Lock mechanism in the financial domain. Furthermore, it demonstrated that training on a single currency yields better results.

Figure 5 compares the overall results of backtesting the baseline model and the proposed model on training data for SC and MC scenarios.

*C. Training time*

The MA-Nolock mechanism significantly outperformed the other two models with its training time in both the SC and MC scenarios (Table VIII, Fig. 6) being the lowest. In comparison, MA-Lock exhibited slightly higher values, in both SC and MC scenarios. SA had the highest values in both scenarios; Therefore, MA-NoLock excels as the most effective model, showcasing its superior performance by a substantial margin, over the other models.

*D. Trading Execution*

Figure 7 demonstrates an snapshot of executed trades in backtesting on the EUR/USD currency pair using MA-lock on the SC scenario. It shows the agent can identify price movements' directions and execute trades, resulting in profitable outcomes.

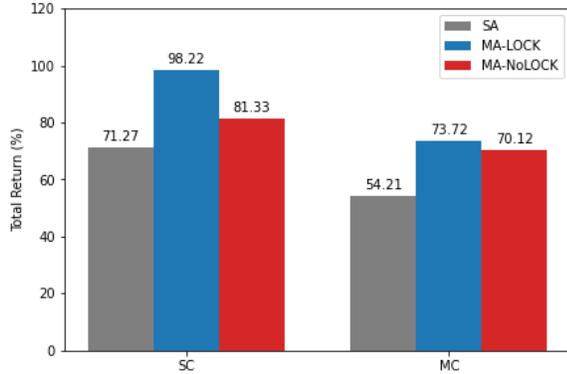

**Fig. 5.** Comparative SC and MC training scenarios backtesting result on EUR/USD.

TABLE VIII
TRAINING TIME IN MINUTES (LOWEST IS BETTER)

| Model | Single currency | Multi-currency |
|---|---|---|
| SA | 347 | 346 |
| MA-Lock | 193 | 251 |
| MA-NoLock | 153 | 218 |

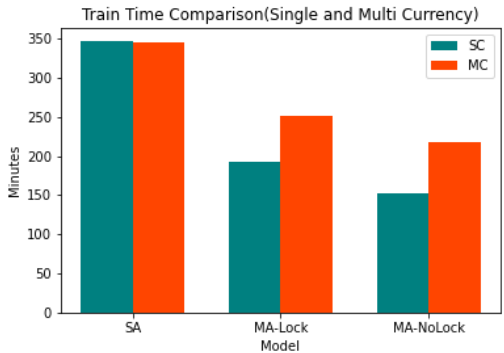

**Fig. 6.** Training time of different models.

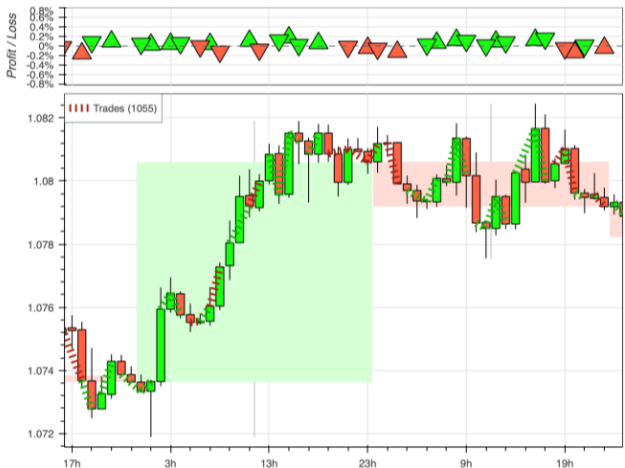

**Fig. 7.** A snapshot of MA-Lock SC scenario backtesting trades on EUR/USD.

## V. CONCLUSION

We aim to train an RL agent capable of trading across diverse assets in forex to make more money while optimizing resource utilization to reduce training time. Our approach involves using the A3C algorithm to distribute training across multiple processes. We explore different training approaches, such as training on single currency pairs and multiple currency pairs, and compare the results. Additionally, we experiment with both single-agent and multi-agent setups, employing PPO as our single-agent algorithm.

Comparing single-agent and multi-agent training, in single currency pair training, both A3C models displayed superior returns and Sharpe Ratio. Multi-currency training showed PPO with generally lower returns, while A3C without Lock stood out with the highest average returns over multiple pairs backtesting and positive Sharpe Ratio. Backtesting results confirmed A3C's superiority over PPO, especially A3C with Lock. Single currency training yielded better overall results. A3C without Lock outperformed other models significantly in training time, asserting its effectiveness in both single and multi-currency scenarios.

Given the complexity of financial markets and the challenge of finding appropriate reward functions for training RL agents, it is recommended for future work to consider the use of No-reward or Reward-Free RL methods.